\documentclass[11pt]{article}
\usepackage{amsmath}
\usepackage{mathrsfs}
\usepackage{amssymb}
\usepackage{titling}
\usepackage{graphicx} 
\usepackage{makeidx}
\usepackage{lipsum}
\allowdisplaybreaks
\usepackage{mathtools}
\usepackage[left=2cm,right=2cm,top=2cm,bottom=2cm]{geometry}
\usepackage{amsfonts}
\usepackage{authblk}
\usepackage[ansinew]{inputenc}
\usepackage[usenames, dvipsnames]{pstricks}
\usepackage{epsfig}
\usepackage{float}
\usepackage{pst-grad} % For gradients
\usepackage{pst-plot} % For axes
\usepackage[colorlinks, hyperindex]{hyperref}
\hypersetup
{
colorlinks, %
citecolor=blue, %
linkcolor=Green, %
urlcolor=blue, %
}

\title{{{\bf Propagation of gravitational waves in various cosmological 
backgrounds}}}

\date{}

\author[1,2]{Sushovan Mondal\thanks{smondal@imsc.res.in}}
\author[1,3]{Saif Ali\thanks{sxa180025@utdallas.edu}}
\author[1]{Shanima S\thanks{ss13ms041@iiserkol.ac.in}}
\author[1]{Narayan Banerjee\thanks{narayan@iiserkol.ac.in}}
\author[1]{Golam Mortuza Hossain\thanks{ghossain@iiserkol.ac.in}}

\affil[1]{Department of Physical Sciences, Indian Institute of Science Education and Research Kolkata, Mohanpur, West Bengal 741246, India.}
\affil[2]{Department of Theoretical Physics, The Institute of Mathematical Sciences, HBNI, Taramani, Chennai, Tamil Nadu 600113, India.}
\affil[3]{Department of Physics, The University of Texas at Dallas, 800 W. Campbell Road, Richardson, TX 75080, USA.}

\begin{document}

\maketitle

\begin{abstract}
The present work investigates some exact solutions of the gravitational wave 
equation in some widely used cosmological spacetimes. The examples are taken 
from spatially flat and closed isotropic models as well as Kasner 
metric which is anisotropic. Various matter distributions are considered, from the standard dust or 
radiation distribution to exotic matters like that with an equation of state $P 
= - \frac{1}{3} \rho$. In almost all cases the frequency and amplitude of the 
wave are found to be decaying with evolution, except for a closed radiation 
universe where there is a resurgence of the waveform, consistent with the 
recollapse of the universe into the big crunch singularity.   

\end{abstract}

%\keywords{}

\section{Introduction}

Albeit the indirect evidence of gravitational waves came through 
the observations by Hulse and Taylor of a binary pulsar \cite{Hulse:1974eb} 
in 1974, the direct detection of the same was possible only very 
recently, when the LIGO and VIRGO scientific collaboration announced their first 
direct observation of gravitational wave in 2016 \cite{Abbott:2016blz}. This 
discovery, not only a great triumph of a relativistic theory of gravitation, but 
also opened up a completely new window for the exploration of the universe. We 
can analyze the properties of gravitational waves to study various extreme 
astrophysical objects like black holes 
\cite{Abbott:2016blz,Abbott:2016nm,Abbott:2017oio}, neutron stars 
\cite{TheLIGOScientific:2017qsa} and also may be able to extract various 
cosmological information \cite{Farr:2019twy} from these gravitational waves.

As some of the detected gravitational waves might have travelled through a 
great distance of a few billion parsecs, such as GW170729 (see the work of 
Abbott {\it et al} \cite{Abbott2019}), the curvature of cosmological background 
might well have its imprint on the waves. The cosmological signature could also 
be small, and might not have any significant observational implications in near 
future, but the estimated distance travelled by them strongly indicates that one 
needs to ascertain the difference of the waveforms travelled through the cosmic 
fluid or through vacuum where the imprints of cosmology is neglected at the 
outset.

No wonder that the investigations along this direction of research is already 
under way. The foundation of such investigations had been in the literature for 
quite a long time back \cite{Isaacson:1967zz}. Mendes and Liddle\cite{liddle}
examined the possible effect of the events in the early universe, such as the early inflation 
or even the possibility of a second burst of inflation at the onset of matter domination, on
the stochastic gravity wave background. Seto and Yokoyama worked on the effect of the equation 
of state of the matter content of the universe in its early stages\cite{seto}. Kuroyanagi, 
Takahashi and Yokoyama\cite{kuro} discussed the features of the early universe by looking at 
the constraints on the primordial gravity waves. Capozziello, Laurentis and Francaviglia considered 
gravity waves in a cosmological background in higher order gravity\cite{capo1}.

 However, the interest naturally 
proliferated after the direct detection of gravitational waves. Very recently 
Valtancoli \cite{valtan} discussed the gravitational wave propagation in an AdS 
spacetime and argued that the graviton mass cannot be measured with  precision 
more than $\sqrt{\Lambda}$. Kulczycki and Malec \cite{malec} showed that for a 
smooth initial metric, with certain differentiability conditions, gives rise to 
gravitational wave pulses that do not interact with the background in the 
radiation dominated era. Possible effects of gravitational waves on the dark 
energy dominated era have been investigated by Creminelli and Vernizzi 
\cite{cremi} and the effect of the waves in connection with structure formation 
was looked at by Jimenez and Heisenberg \cite{heisen}. Caldwell and Devulder 
worked on the possible opacity for gravitational waves by a particular form of 
dark energy and showed that the maximum absorption is around $1\%$ for the 
redshift range $0.5\leq z \leq 1.5$ \cite{cald}.  Most of these
investigations are based on spatially flat isotropic cosmologial background.
Bradley, Forsberg 
and Keresztes \cite{brad} discussed the propagation of gravitational waves in a 
class of locally rotationally symmetric spacetime. There are some work on the 
propagation of such waves in relativistic theories of gravity other than general 
relativity also, such as in Jordan-Brans-Dicke theory\cite{dunya}, in teleparallel 
gravity\cite{rafael1}, in a general scalar-tensor theory\cite{rafael2}. Evolution of 
gravitons in an accelerating in an extended gravity theory was discussed by 
Capozziello {\it et al} \cite{capo2}. 

The motivation of the present work is to look at the wave forms travelling in a 
cosmological background, through the cosmic fluid.  We do not bother about source of 
the wave, but rather concentrate on the exact solutions of the wave equation in the 
linear regime of the perturbation. But the 
investigation is not restricted to a spatially flat isotropic cosmology only.  
From the general equation 
for the gravitational wave, we write down the equations for some given 
cosmology, namely a spatially flat FRW cosmology, a spatially closed FRW 
cosmology and an anisotropic cosmology. We neglect the perturbation in the 
matter sector, and consider only the metric perturbation giving rise to 
perturbations in the curvature. In some well known examples of cosmological 
models, we could actually find the exact solutions to the wave equation. For 
some choices of the constants of integrations, we plot the wave form. The 
interest in this work is to look at the features qualitatively, so actual 
numbers related to various parameters are chosen for the sake of convenience so 
as to get exact solutions. As expected, in most cases both the amplitude and 
frequency die down with time, with one exception of a spatially closed model 
where the wave form has a resurgence with the recollapse of the universe.

The closest to this present work is the one by Flauger and Weinberg 
\cite{wein}, which deals with the propagation in a cold dark matter (CDM). They 
predict that the magnitude of the cosmic effect is too small to influence the 
observations. This is attributed by them to the fact that the CDM is too 
nonrelativistic to affect the wave. They also show that for a primordial wave, 
the imprints are more pronounced for the intermediate modes. We, however, looked 
at various kinds of matter distribution including very highly relativistic 
matter like a radiation distribution.

The outline of the paper is as follows. We begin with a brief review of the 
general formulation involved in the propagation of gravitational wave in Sec. 
II and write down the wave equation of the gravitational wave in a general 
curved background. In Sec. III, we consider the spatially homogeneous and 
isotropic universe. The equations for the gravitational wave for the FRW 
universe are written and exact solutions are obtained for spatially flat and 
spatially closed FRW universe for some given matter distributions. In Sec. IV 
we discuss about the wave propagation in an anisotropic universe by considering 
Bianchi universe. The exact solutions for a particular orientation of 
propagation are obtained for a Kasner metric. A discussion of the results 
obtained are given in the fifth and final section.

\section{The General Formulation}

We begin by writing the equation for the metric perturbations in a general 
curved spacetime whose solution will finally yield the gravitational wave 
profile for the given  spacetime. The procedure is standard and is discussed  in 
standard texts like \cite{Padmanabhan:2010zzb}. One can also refer to 
\cite{Svitek:2006yj,Ford:1977dj}. We shall consider a non-zero background 
stress-energy tensor. The perturbed metric is written as
\begin{equation}
g_{\mu \nu}= \gamma_{\mu \nu}+\epsilon h_{\mu \nu}, \label{background}
\end{equation}
where $\epsilon$ is a small parameter which makes $\epsilon h_{\mu \nu}$ a 
small perturbation to $\gamma_{\mu \nu}$, which is the background metric. 
Raising and lowering of indices will take place through the background  metric 
$\gamma_{\mu \nu}$. The inverse of the metric, i.e.,  $g^{\mu\nu}$, can be 
written as (up to linear order)
\begin{equation}
	g^{\mu \nu}= \gamma^{\mu \nu}-\epsilon h^{\mu \nu}. \label{backgroundinv}
\end{equation}

Unlike Minkowski background, where the perturbation is considered to be small 
compared to unity, the isolation of the perturbation from the background is not so trivial.
This is primarily because both $\gamma_{\mu \nu}$ and $h_{\mu \nu}$ are functions of 
coordinates. What one normally considers in this case a short wave or high frequency 
condition where the wavelength ($\lambda$)  of the perturbation, which satisfies a wave equation, 
is much smaller compared to the typical length scale ($L$) of the backgroumd. It can be shown that 
the first order correction to the Christoffel symbol $\Gamma$ (called $\Gamma^{(1)}$ in the 
present discussion) and the first order correction to the Ricci curvature ($R^{(1)}$) corresponds to 
the short wave. For an elaborate calculation, we refer to the comprehensive book by 
Maggiore\cite{Maggiore:1900zz}. 

Using the equations (\ref{background}) and (\ref{backgroundinv}) in the 
expression of the Christoffel symbol, one can obtain 
\begin{align}
	\Gamma^\rho_{\mu \nu}=&\overline{\Gamma}^\rho_{\mu 
\nu}+\frac{1}{2}\gamma^{\rho \sigma}[\nabla_\nu h_{\mu \sigma}+\nabla_\mu 
h_{\sigma \nu}-\nabla_\sigma h_{\mu\nu }]\\=&\overline{\Gamma}^\rho_{\mu 
\nu}+\Gamma^{\rho (1)}_{\mu \nu}. \label{christ}
\end{align} 
We will deal with only the terms that are linear in $\epsilon$ and hence the 
use of term $\epsilon$ is absorbed in the components $h^{\mu \nu}$ which 
themselves will take care of being small. We can clearly see that in the case of 
curved background spacetime, the Christoffel symbols break down into two 
distinct parts, one is the background spacetime Christoffel symbols which are 
denoted by the  $\overline{\Gamma}$ and the another one is linearly perturbed 
Christoffel symbols which are denoted as $\Gamma^{(1)}$. One can use the 
Christoffel symbols to obtain Riemann tensor. We write the Riemann tensor as a 
sum  of the background Riemann tensor and the linearly perturbed Riemann tensor 
as 
\begin{equation}
	{R^\rho}_{ \mu \nu \lambda}= {\overline{R}^\rho}_{ \mu \nu 
\lambda}+{R^{(1)\rho}}_{\mu \nu \lambda}. \label{riemann}
\end{equation}
The linearly perturbed part of Ricci tensor can now be written as
\begin{align}
R^{(1)}_{\alpha \beta}=
\frac{1}{2}\left[\nabla_\rho \nabla_\alpha h^\rho_\beta  +\nabla_\rho 
\nabla_\beta h^\rho_\alpha -\Box h_{\alpha\beta}-\nabla_\beta \nabla_\alpha h 
\right]. \label{riccipert}
\end{align}
In the above equation $h=h^\mu_\mu$ is the trace of the metric perturbation.

In a similar fashion we can divide Einstein tensor, and hence the Einstein 
field equation, into its background part $\overline{G}_{\alpha\beta}$ and the 
linearly perturbed part $G^{(1)}_{\alpha\beta}$ as follows 
\begin{align}
G_{\alpha\beta} &= \overline{G}_{\alpha\beta}+ 
G^{(1)}_{\alpha\beta}\\&=-8\pi T_{\alpha\beta}\\&= -8\pi 
\left(\overline{T}_{\alpha\beta}+ T_{\alpha\beta}^{(1)}\right), 
\label{einsteintensor}
\end{align}
where the $\overline{T}_{\alpha\beta}$ is the background stress-energy 
tensor and $T_{\alpha\beta}^{(1)}$ is the linear order perturbations of 
stress-energy tensor. From the above discussion of order of magnitude, we 
consider the perturbation of stress-energy tensor to be zero, i.e. 
$T^{(1)}_{\alpha\beta}=0$. So one has  $G^{(1)}_{\alpha\beta}=0$. 

This requires some explanation. As shown in \cite{Maggiore:1900zz}, the first order 
correction to the stress energy tensor actually contributes to long wave (or small 
frequency) and not to the short wave condition ($\lambda << L$) and thus not of 
relevance for the present discussion as discussed in the beginning of this section.

If we take the trace of Einstein equation, we get $R^{(1)} = -8\pi T^{(1)}=0$ and hence, we get
$R^{(1)}_{\alpha\beta} = 0$. Therefore, the expression of 
$R^{(1)}_{\alpha\beta}$ (\ref{riccipert}), leads to
\begin{align}\label{RicciPerturbedEq}
R^{(1)}_{\alpha \beta}=\frac{1}{2}\left[\nabla_\rho \nabla_\alpha 
h^\rho_\beta  +\nabla_\rho \nabla_\beta h^\rho_\alpha -\Box 
h_{\alpha\beta}-\nabla_\beta \nabla_\alpha h \right]= 0.
\end{align}
The above equation can be simplified using the {\it trace-reversed} 
perturbation variable. The {\it trace-reversed} perturbation variable is defined 
as $\tilde{h}_{\mu \nu}=h_{\mu \nu}-\frac{1}{2}\gamma_{\mu \nu}h.$ As 
$\tilde{h}=-h$, where $\tilde{h}= \tilde{h}^\mu_\mu$ and $h= h^\mu_\mu$, we 
have  $h_{\mu \nu}=\tilde{h}_{\mu \nu}-\frac{1}{2} \gamma_{\mu \nu} \tilde{h}.$ 
One can replace the perturbation variable with the {\it  trace-reversed} 
perturbation variable, as a result, the equation (\ref{RicciPerturbedEq}) 
reduces to
% 
% %\onecolumngrid
% \begin{align}
% R^{(1)}_{\alpha \beta}= \nabla_\rho \nabla_\alpha 
% \tilde{h}^\rho_\beta+\frac{1}{2} \nabla_\beta \nabla_\alpha 
% \tilde{h}-\frac{1}{2}  \nabla_\alpha \nabla_\beta \tilde{h} 
% +\nabla_\rho \nabla_\beta \tilde{h}^\rho_\alpha \nonumber\\
% -\Box \tilde{h}_{\alpha \beta}+\frac{1}{2}\gamma_{\alpha\beta}\Box 
% \tilde{h}=0. 
% \end{align}
% %\end{widetext}
% \twocolumngrid
%
\begin{equation}
\nabla_\rho \nabla_\alpha 
\tilde{h}^\rho_\beta+\nabla_\rho \nabla_\beta \tilde{h}^\rho_\alpha-\Box 
\tilde{h}_{\alpha \beta}+\frac{1}{2}\gamma_{\alpha\beta}\Box \tilde{h}=0. 
\label{riccipertnew}
\end{equation}
The equation (\ref{riccipertnew}) can be further simplified by using a suitable 
gauge. In the case of curved background spacetime, the modified Lorenz gauge can 
be written as 
\begin{equation}
\nabla_\nu \tilde{h}^{\mu \nu}=0. \label{lorgauge}
\end{equation}

 It may be noted that the corresponding equation given in \cite{kuro} looks different from 
this equation, which is simply because the present work uses the cosmic time $t$ as opposed to the 
conformal time $\tau$ used in \cite{kuro}. A transformation of the form $dt = a(\tau) d\tau$ in the present 
equation resolves the apparent mismatch.  
We can modify the first and second terms in the equation (\ref{riccipertnew}) 
in terms of Riemann and Ricci tensors. Using Lorenz gauge, equation 
(\ref{riccipertnew}) reduces to
\begin{equation}
 \bar{R}_{\delta \alpha} \tilde{h}^\delta_\beta+\bar{R}_{\delta \beta} 
\tilde{h}^\delta_\alpha-2\bar{R}_{\beta \delta \alpha \rho} \tilde{h}^{\delta 
\rho}-\Box \tilde{h}_{\alpha \beta}+\frac{1}{2}\gamma_{\alpha\beta}\Box 
\tilde{h}=0. \label{gravwaveeqn}
\end{equation}
Apart from the Lorenz gauge, we also impose the condition that gravitational 
wave is traceless, and therefore we have $\tilde{h}=\tilde{h}^\alpha_\alpha=0$. 
Using the traceless condition in equation (\ref{gravwaveeqn}), we obtain the 
final wave equation for the gravitational wave in the curved background 
spacetime as
\begin{equation}
 \Box \tilde{h}_{\beta \alpha}-2\bar{R}_{\delta \beta \alpha \rho} 
\tilde{h}^{\delta \rho}-\bar{R}_{\delta \alpha} 
\tilde{h}^\delta_\beta-\bar{R}_{\delta \beta} 
\tilde{h}^\delta_\alpha=0.\label{curgravwaveequation}
 \end{equation} 
It may be mentioned that in the equations (\ref{gravwaveeqn}, \ref{curgravwaveequation}), the effect of the 
matter distribution on the gravitational wave propagation is taken 
care of by the Ricci tensor through the background Einstein equations, while 
the Riemann tensor determines the contribution due to the background  
curvature.

\section{Gravitational waves in FRW universe}

In this section we investigate the propagation of gravitational waves in a 
cosmological background which is spatially homogeneous and isotropic. The 
corresponding spacetime is described by the Friedmann, Robertson and Walker 
(FRW) metric 
\begin{equation}
ds^2 = -dt^2 +a^2(t)\left[\frac{dr^2}{1-k r^2}+r^2 
d\theta^2+r^2\sin^2\theta d\phi^2\right] ,\label{met} 
\end{equation}
where $k$ denotes the spatial curvature and takes values $0,+1,-1$ for 
spatially flat, closed and open universe respectively.

\subsection{Background equations}

The function $a(t)$ is known as the scale factor and is governed, when the 
universe is filled with a perfect fluid, by the equations
\begin{align}
\frac{\ddot{a}}{a} &= -\frac{4\pi}{3}(\rho+3p)\\
\left(\frac{\dot{a}}{a}\right)^2 &= \frac{8\pi}{3}\rho -\frac{k}{a^2} 
.\label{fried}
\end{align}
For a given equation of state $P = P(\rho)$, these equations can be 
integrated to find the dynamics of the background metric.   

\subsection{Gravitational wave equations}

To derive gravitational wave equations in FRW spacetime we use the equation 
(\ref{curgravwaveequation}) along with the relevant Riemann and Ricci tensor 
components which are computed using the FRW metric (\ref{met}). For brevity of 
notations, now onward we shall use $h_{\alpha\beta}$ instead of 
$\tilde{h}_{\alpha\beta}$ to denote metric perturbation components. In 
particular, the gravitational wave equations in spherical polar coordinates, can be explicitly expressed as 
\allowdisplaybreaks
%
%\begin{widetext}
%\onecolumngrid
\begin{eqnarray}
%\begin{center}
&\Box {h}_{00} - 
2\left[\frac{\ddot{a}(1-kr^2)}{a^3}{h}_{11}+\frac{\ddot{a}}{a^3r^2}{h}_{22}
+\frac{\ddot{a}}{a^3r^2\sin^2{\theta}}{h}_{33}\right]-6\frac{\ddot{a}}{a}{h}_{00
} = 0,\\ \label{eq1}
&\Box {h}_{11} - 
2\left[\frac{\dot{a}^2+k}{a^2r^2(kr^2-1)}{h}_{22}+\frac{\dot{a}^2+k}{
(kr^2-1)a^2r^2\sin^2\theta}{h}_{33}-\frac{a\ddot{a}}{kr^2-1}{h}_{00}\right]
-2\left[ 
2\frac{\dot{a}^2}{a^2}+2\frac{k}{a^2}+\frac{\ddot{a}}{a}\right]{h}_{11}=0,\\ 
\label{eq2}
&\Box{h}_{22}+ 
2\left[\frac{r^2(\dot{a}^2+k)(1-kr^2)}{a^2}{h}_{11}+\frac{(\dot{a}^2+k)}{
a^2\sin^2\theta}{h}_{33} - r^2 a \ddot{a}{h}_{00} 
\right]-2\left[2\frac{\dot{a}^2}{a^2}+2\frac{k}{a^2}+\frac{\ddot{a}}{a}\right]{h
}_{22} = 0,\\ \label{eq3}
&\Box {h}_{33}- 2\left[r^2a\ddot{a}\sin^2\theta{h}_{00} 
-\frac{(1-kr^2)r^2\sin^2\theta(\dot{a}^2+k)}{a^2}{h}_{11}-\frac{
\sin^2\theta(\dot{a}^2+k)}{a^2}{h}_{22}\right]-2\left[2\frac{\dot{a}^2}{a^2}+2\frac{k}{a^2}+\frac{\ddot{a}}{a}\right]{h
}_{33} = 0, \\ \label{eq4}
&\Box {h}_{0i}- 
\left[6\frac{\ddot{a}}{a}+2\frac{\dot{a}^2}{a^2}+\frac{2k}{a^2}\right]{h}_{0i} 
= 0, \text{ for}\ i = 1,2,3 ,\\ \label{eq5}
&\Box 
{h}_{mn}-\left[2\frac{\ddot{a}}{a}+6\frac{\dot{a}^2}{a^2}+\frac{6k}{a^2}\right]{
h}_{mn} = 0, \text{ for}\ m,n = 1,2,3 \text{ and}\ m \neq n. \label{eq8}
%\end{center}
\end{eqnarray}
%\end{widetext}
%\twocolumngrid

%\changed{GMH: Above equations can be written in a compact way. For eg. use 
%$h_{0i}$ for $i=1,2,3$ etc.}

Here $x^{1}, x^{2}, x^{3}$ correspond to the coordinates $r, \theta, \phi$ respectively.

\subsection{Solutions to the gravitational wave equations}

\subsubsection{Spatially flat FRW universe}

In spatially flat FRW spacetime, it is convenient to use Cartesian coordinate 
system, which allows us use transverse-traceless-synchronous (TTS) 
gauge \cite{Grishchuk:1981bt,Caldwell:1993xw} efficiently. We choose that the 
wave is propagating along the $x$ direction. So according to TTS gauge, 
$h_{0\mu}$ and ${h}_{1\mu}$ are zero, and ${h}_{22} = -{h}_{33}$, where the 
subscript $0,1,2,3$ denotes $t,x,y,z$ coordinates respectively. After applying 
the TTS gauge, the number of nonzero and independent wave equations reduce to 
two, which are given by 
\begin{eqnarray}
\Box {h}_{22}-\left[2\frac{\ddot{a}}{a}+6\frac{\dot{a}^2}{a^2}\right]{h}_{22} 
&=& 0 \label{flat1}\\
\Box {h}_{23}-\left[2\frac{\ddot{a}}{a}+6\frac{\dot{a}^2}{a^2}\right]{h}_{23} 
&=& 0\label{flat2}.
\end{eqnarray}
We shall investigate the evolution of gravitational waves in a dust  
dominated as well as radiation dominated universe.

\begin{figure}[H]
    \centering
    \includegraphics[width=0.7\linewidth]{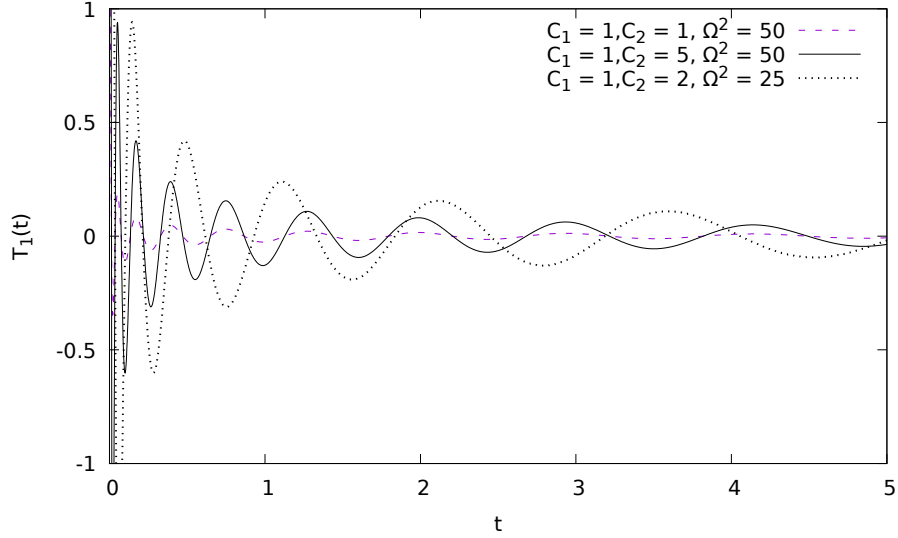}
    \caption{Plot of evolution of gravitational wave with time in a dust matter 
dominated spatially flat universe.} %The time axis is plotted using arbitrary 
%unit.}
    \label{fig:dust}
\end{figure}

For the dust ($P=0$) dominated spatially flat FRW universe, the scale 
function grows with time as
\begin{equation}
    a = a_i~t^{\frac{2}{3}} ~,\label{scaledust}
\end{equation}
where $a_i$ is a constant. 
Now we consider an ansatz that the perturbation components are function of $t$ 
and $x$ only. As equations (\ref{flat1}) and (\ref{flat2})  are similar in 
form, so it is sufficient to solve one of them. We take ${h}_{22} = 
\Psi(x,t)$.  Equation (\ref{flat1}) will now look like 
\begin{equation}
    \frac{\partial^2 \Psi}{\partial t^2} + 3\frac{\dot{a}}{a}\frac{\partial 
\Psi}{\partial t} - \frac{1}{a^2}\frac{\partial^2 \Psi}{\partial x^2} - 
\left(2\frac{\ddot{a}}{a}+6\frac{\dot{a}^2}{a^2}\right)\Psi = 0. \label{flateqn}
\end{equation}
For dust dominated universe, we can use the expression of scale factor from 
equation (\ref{scaledust}), in equation (\ref{flateqn}) to get the wave 
equation as
\begin{equation}
    \frac{\partial^2 \Psi}{\partial t^2} +\frac{2}{t} \Psi -\frac{1}{a_i^2 
t^{\frac{4}{3}}} \frac{\partial^2 \Psi}{\partial x^2} - \frac{20}{9t^2} \Psi = 
0.\label{flatdust}
\end{equation}
One can solve this equation by using separation of variable as $\Psi(x,t) = 
X_1(x)T_1(t)$. The equation (\ref{flatdust}) will break as                  \begin{equation}
a_i^2 t^{\frac{4}{3}}\left(\frac{1}{T_1}\frac{d^2 T_1}{d t^2} + 
\frac{2}{t} \frac{1}{T_1} \frac{d T_1}{d t} - \frac{20}{9t^2}\right) = 
\frac{1}{X_1}\frac{d^2 X_1}{d x^2} = -{\Omega^2} ~, \label{flatsep}
\end{equation}
where ${\Omega^2}$ is the separation constant.The choice of the sign of the separation 
constant is negative as we are looking for a periodic variation. From equation (\ref{flatsep}), 
we can see that the solution of the spatial part is given by 
\begin{equation}
    X_1(x) = E \cos(\Omega x) + F \sin(\Omega x), \label{spatial}
\end{equation}
where $E$ and $F$ are constant of integration. The temporal part of the 
equation is 
\begin{equation}
    \frac{d^2 T_1}{d t^2} + \frac{2}{t} \frac{d T_1}{d t} + 
\left(\frac{{\Omega^2}}{a_i^2 t^{\frac{4}{3}}} - \frac{20}{9t^2} \right) T_1 = 0.
\end{equation}
The solution of the above equation is 
\begin{equation}
\begin{split}
T_1(t) = \frac{2 \sqrt{\frac{2}{3}} 
a_i^{\frac{3}{2}}}{3\sqrt{t}{\Omega}^{\frac{3}{2}}}\left[C_1 \Gamma 
\left(1-\frac{\sqrt{89}}{2}\right) J_{-\frac{\sqrt{89}}{2}}\left(\frac{3 
\sqrt[3]{t} {\Omega} }{a_i}\right)\right. \\
+ \left.C_2 \Gamma \left(1 + \frac{\sqrt{89}}{2}\right) 
J_{\frac{\sqrt{89}}{2}}\left(\frac{3 \sqrt[3]{t} {\Omega} 
}{a_i}\right)\right].
\end{split}
\end{equation}
Here $C_1$ and $C_2$ are the constants of integration, $J_n(z)$ and $\Gamma(z)$ 
are Bessel function of first kind of order $n$ and gamma function respectively. 
The plot of the evolution of gravitational wave with time is given by 
FIG. \ref{fig:dust}, where the constant $a_i$ is taken as $1$ without any loss 
of generality. The unit of time needs some attention. In the metric (\ref{met}), as 
$k$ is dimensionless, the radial coordinate $r$ is scaled by some $r_0$. The time is 
actually scaled by some $t_0$, so that all the quantities in the metric are dimensionless, 
which is quite a standard practice. From equation (\ref{scaledust}), one can ascertain 
$t_0 = a_i$, which is chosen to be unity. With this, one can see that the unit of time is
chosen to be $H_0^{-1}$ multiplied by a constant of order unity (such as $\frac{2}{3}$ for dust  
and $\frac{1}{2}$ for radiation etc). $H_0$ is the Hubble parameter $H=\frac{\dot{a}}{a}$ evaluated at 
$a=a_i=1$. In all the plots, the amplitude is scaled by its initial value.

From the plot we see that the amplitude of the wave decays and the frequency of 
the wave also decreases with time. One may understand that the damping is due 
to the expansion of the universe, which describes the behaviour of the 
gravitational waves in an expanding universe with respect to a co-moving 
observer.  

For a radiation dominated universe, with the equation of state being $P = 
\frac{1}{3} \rho$, the scale function grows as
\begin{equation}
    a = a_i t^{\frac{1}{2}}.\label{scaleradiation}
\end{equation}

As earlier, by choosing $a_i = 1$ and using the separation of variables ${\Psi}_{2} = X_2 (x) T_2 (t)$, one 
may note that the spatial part is the same as the dust dominated universe, and 
the solution is given by the equation (\ref{spatial}). The temporal evolution 
equation of the gravitational wave is given by
\begin{equation}
    \frac{d^2 T_2}{d t^2} + \frac{3}{2t} \frac{d T_2}{d t} + 
\left(\frac{{\Omega^2}}{a_i^2 t} - \frac{1}{t^2} \right) T_2 = 0 ~,
\end{equation}
and whose solutions are of the form
\begin{equation}
\begin{split}
T_2(t) = \frac{\sqrt{a_i}}{({\Omega^2} t)^\frac{1}{4}}\left[C_3 \Gamma 
\left(1-\frac{\sqrt{17}}{2}\right) J_{-\frac{\sqrt{17}}{2}}\left(\frac{2 
\sqrt{t} {\Omega} }{a_i}\right)\right.\\
+ \left. C_4 \Gamma \left(1+\frac{\sqrt{17}}{2}\right) 
J_{\frac{\sqrt{17}}{2}}\left(\frac{2 \sqrt{t} 
{\Omega} }{a_i}\right)\right] \label{flatradtemp},
\end{split}
\end{equation}
where $C_3$ and $C_4$ are constants of integration.
\begin{figure}[H]
\centering
\includegraphics[width=0.7\linewidth]{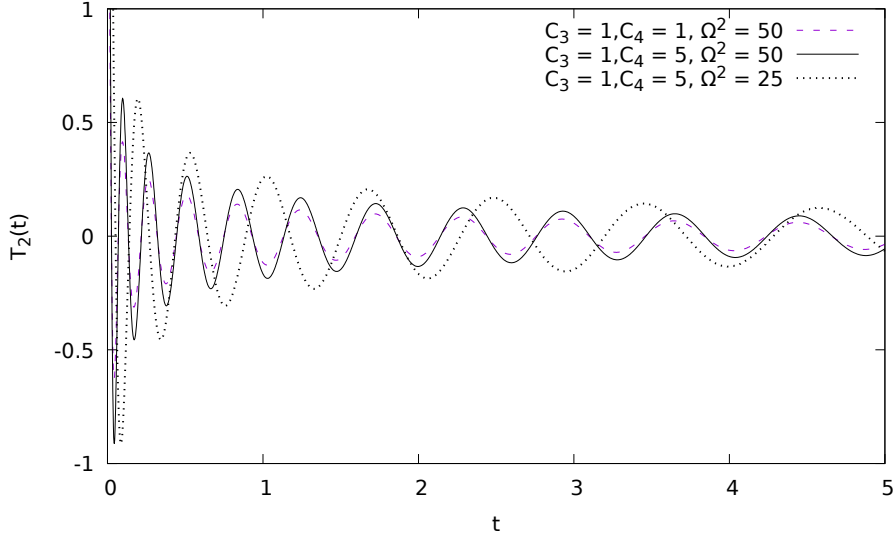}
\caption{Plot of evolution of gravitational wave with time in a radiation 
dominated spatially flat universe.} %The time axis is plotted using arbitrary 
%unit.}
\label{fig:radflat}
\end{figure}
Similar to the evolution of the waveform in a dust dominated universe,
here the amplitude and frequency of the wave decreases with time. From FIG. 
\ref{fig:radflat}, it is quite apparent that the rate of damping of the wave is 
higher in dust dominated universe than the radiation dominated universe. This is 
actually consistent as the expansion is faster in the dust dominated universe 
compared to the radiation dominated universe.

\subsubsection{Spatially closed FRW universe}

In spatially closed FRW universe, we can again make use of the gauge freedom to 
reduce the number of the independent equations. For spatially closed universe, system of  
spherical polar coordinates is the most suitable one to analyze various properties 
of the universe. We take that the wave propagates in 
radial direction. We also consider the observer is comoving, so the 4-velocity 
of the observer is $U^\mu = (1,0,0,0)$. In TTS gauge 
\cite{Grishchuk:1981bt,Caldwell:1993xw} the equations  of independent nonzero 
perturbation components are 
\begin{eqnarray}
\Box 
{h}_{22}-\left[2\frac{\ddot{a}}{a}+6\frac{\dot{a}^2}{a^2}+\frac{6}{a^2}\right]{h
}_{22} &=& 0 \label{fin1}\\
\Box 
{h}_{23}-\left[2\frac{\ddot{a}}{a}+6\frac{\dot{a}^2}{a^2}+\frac{6}{a^2}\right]{h
}_{23} &=& 0\label{fin2}.
\end{eqnarray}
Here the subscript $2$ , $3$ denotes $\theta$ and $\phi$ coordinates 
respectively in spherical polar coordinate. 

We shall investigate the solutions of the gravitational wave equation for two 
different kinds of evolution, namely the radiation dominated decelerated 
universe and a coasting universe \cite{Kolb:1989bg}. For radiation (equation of 
state $P/\rho = 1/3$) dominated spatially closed universe the scale function 
evolves as 
\begin{equation}
    a(t) = \zeta \sqrt{1 - \left(1 - \frac{t}{\zeta}\right)^2}.\label{radsc}
\end{equation}
Here $\zeta = \sqrt{2q_0/(2q_0 - 1)}$ is constant and $q = -\ddot{a}/a H^2$, is the 
deceleration parameter of universe and a suffix $0$ indicates its present 
value. From the equation (\ref{radsc}), it is understood that the scale factor 
grows at a decreasing rate until it stops and gives way to a contraction. 

Other than radiation, we shall also discuss about a matter whose equation of 
state is given by $P/\rho = -1/3$. In various literature this type of matter is 
described as K-matter \cite{Kolb:1989bg} or winding modes of a ``string gas'' 
\cite{Gasperini:2007zz}. For this type of a matter distribution, the scale 
factor evolves in terms of cosmic time as \cite{Kolb:1989bg}
\begin{equation}
    a(t) = \sqrt{(K-1)} t = a_0 H_0 t. \label{ksc}
\end{equation}
Here $K = 8\pi G{\rho}_0  {a_0}^2/3$, and $a_0$ and $H_0$ are scale factor and 
Hubble parameter at present time respectively. This evolution has the deceleration 
parameter $q=0$, which justifies the name a ``coasting universe''. From various observational 
parameters one can constrain the value of $K$ as $K\geq 1.2$ 
\cite{Kolb:1989bg}. 

To solve the gravitational wave equations, we assume that perturbation 
components are the functions of time($t$) and radial distance($r$) only. As the 
equations (\ref{fin1}) and (\ref{fin2}) are same, so it is sufficient to 
consider only one component of the metric perturbations, namely $h_{22} = 
\Phi$. With the assumption $\Phi = \Phi(r, t)$, the equation (\ref{fin1}) reads 
as
\begin{equation}
\begin{split}
-\frac{\partial^2 \Phi}{\partial t^2} -3 \frac{\dot{a}}{a}\frac{\partial 
\Phi}{\partial t} + \frac{(1- r^2)}{a^2} \frac{\partial^2 \Phi}{\partial r^2} + 
\frac{(2-3r^2)}{a^2 r}\frac{\partial \Phi}{\partial r}\\
- \left[2\frac{\ddot{a}}{a}+6\frac{\dot{a}^2}{a^2}+\frac{6}{a^2}\right]\Phi = 0 
. \label{ansatz}
\end{split}
\end{equation}
We can solve this equation by the method of separation of variables, $\Phi(r,t) 
= R_3(r)T_3(t)$. The possible solutions of radial part is discussed in 
literature \cite{DEath:1976dwo,Bicak:1996qw}. The temporal evolution of the wave 
is given by 
\begin{equation}
    \frac{d^2 T_3}{d t^2} + 3 \frac{\dot{a}}{a} \frac{d T_3}{d t} + 
\left[2\frac{\ddot{a}}{a}+6\frac{\dot{a}^2}{a^2}+\frac{(6-{\Omega^2})}{a^2}
\right] T_3 = 0 ~, \label{temp}
\end{equation}
where ${\Omega^2}$ is the separation constant.

\begin{figure}[H]
    \centering
    \includegraphics[width=0.7\linewidth]{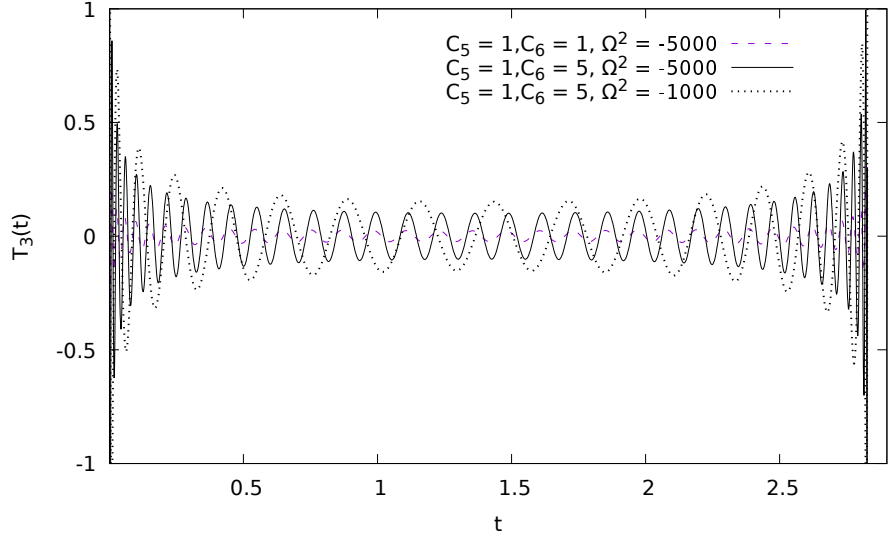}
    \caption{Plot of evolution of gravitational wave with time in a radiation 
dominated spatially closed universe.} %The time axis is plotted using arbitrary 
%unit.}
    \label{fig:radclose}
\end{figure}

\begin{figure}[H]
    \centering
    \includegraphics[width=0.7\linewidth]{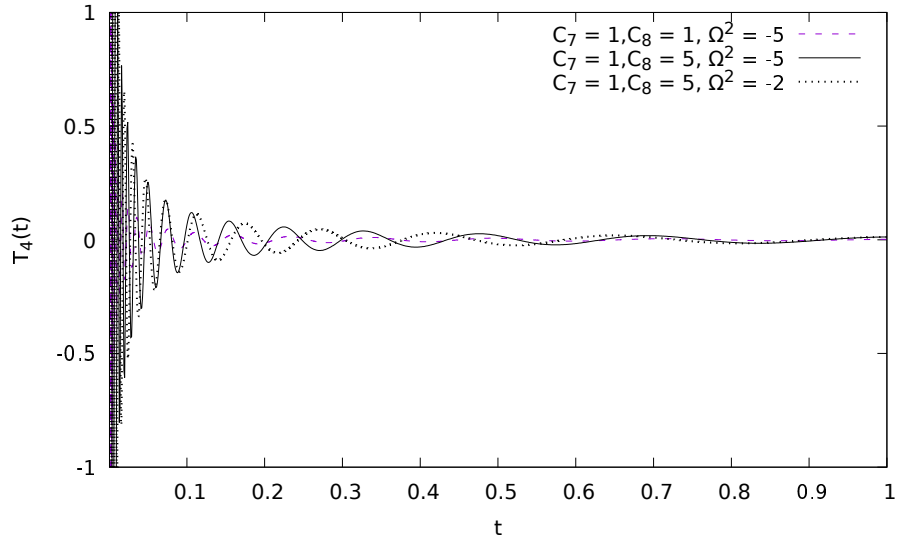}
    \caption{Plot of evolution of gravitational wave with time in a K-matter 
dominated spatially closed universe.}% The time axis is plotted using arbitrary 
%unit.}
    \label{fig:coasting}
\end{figure}

For radiation dominated universe the scale factor is given by equation 
(\ref{radsc}). Using equation (\ref{radsc}) and (\ref{temp}), we can get the 
solution of temporal evolution of GW in radiation dominated universe, which is 
given by
\begin{equation}
  T_3(t) = \frac{1}{{\sqrt[4]{t} \sqrt[4]{t-2\zeta}}}\left[
C_5P^\alpha_\beta\left(\frac{t}{\zeta}-1\right) + 
C_6 Q^\alpha_\beta\left(\frac{t}{\zeta}-1\right) \right] ~, \label{tempfin}
\end{equation}
where $\alpha = (i\sqrt{15}/2)$ and $\beta = ((2\sqrt{1-\Omega^2} - 1)/2)$. 
$P_{a}^{b}(x)$ and $Q_{a}^{b}(x)$ are associated Legendre functions of the first 
and the second kind respectively. $C_5$ and $C_6$ are constants of integration. 
We choose the deceleration parameter at present time to be unity to plot 
the function as given in FIG. \ref{fig:radclose}. We find that both amplitude 
and frequency of the wave decreases during the expanding period of the universe 
and increases later when the universe starts to collapse. We also note another 
important feature that the amplitude blows up at the end of the evolution which 
corresponds to the collapse of the universe to a singularity.  
Now we  discuss about the evolution of gravitational wave in a K-matter 
dominated spatially closed universe.  We have already seen that the universe 
expands with a $q=0$ for K-matter dominated spatially closed 
universe. Using equation (\ref{ksc}) and (\ref{temp}), we can get the solution 
of evolution of time dependent part, $T_4(t)$, for K-matter dominated universe, 
which is given by

\begin{equation}
  T_4(t) =   \frac{1}{t}\left(C_7 t^{-\sigma} + C_8 t^\sigma\right) ~,
\end{equation}
where $\sigma = ( \sqrt{\Omega^2 - 5K - 1}/ \sqrt{K-1})$ and $C_7$, $C_8$ are 
constants of integration. 
We plot $T_4(t)$ with respect to time $t$ by choosing numerical value of $ K = 
1.2$ \cite{Kolb:1989bg} and is given in FIG. \ref{fig:coasting}.
From this plot one can see that the amplitude of the wave is decaying and the 
frequency is also decreasing with time. A resurgence of the amplitude and 
frequency of the gravitational wave is not indicated in this case. This is 
quite expected as the universe does not re-collapse.

\section{Gravitational waves in Bianchi type-I Universe}

\subsection{Bianchi-I metric}

The observed universe, to a very high degree of accuracy ($1$ in $10^{-5}$) is 
isotropic. However, an anisotropic universe warrants discussion for various 
purposes (see the work of Rodrigues \cite{rodrig}). In this section, a Bianchi 
type-I cosmological model is taken as the background spacetime for the 
propagation of gravitational waves. A Bianchi type-I universe, the simplest 
homogeneous but anisotropic cosmological model, is described  by the metric
\begin{equation} \label{b1metric}
ds^2=-dt^2+a^2(t) dx^2+b^2(t) dy^2+c^2(t) dz^2,
\end{equation}
where $a,b,c$ are functions of the cosmic time $t$. These functions can be 
viewed as the scale factors along the three spatial axes $x,y,z$ respectively.

\subsection{Gravitational wave Equations}

One can write down the non-zero components of the Christoffel symbol, Riemann 
tensor and Ricci tensor using the Bianchi type-I metric (\ref{b1metric})
as needed in the general gravitational wave equation 
(\ref{curgravwaveequation}). The individual wave equation for the $10$ 
components of the metric perturbation $h_{\alpha \beta}$ can be explicitly 
written as  
\allowdisplaybreaks
%\begin{widetext}
\begin{eqnarray}
	 {\Box h_{00}-2\left(\left(\frac{\ddot{a}}{a^3}\right) 
h_{11}+\left(\frac{\ddot{b}}{b^3}\right) 
h_{22}+\left(\frac{\ddot{c}}{c^3}\right) h_{33} \right)-2\left( 
\frac{\ddot{a}}{a}+\frac{\ddot{b}}{b}+\frac{\ddot{c}}{c}  \right) h_{00}=0,}\\
	{\Box h_{11}+2\left(\left(\frac{a\dot{a}\dot{b}}{b^3}\right) 
h_{22}+\left(\frac{a\dot{a}\dot{c}}{c^3}\right) h_{33}-\left(a\ddot{a}\right) 
h_{00} \right)-\frac{2}{a}\left( 
\ddot{a}+\frac{\dot{a}\dot{b}}{b}+\frac{\dot{a}\dot{c}}{c}  \right) h_{11}=0,}\\
	{\Box h_{22}+2\left(\left(\frac{b\dot{a}\dot{b}}{b^3}\right) 
h_{11}+\left(\frac{b\dot{b}\dot{c}}{c^3}\right) h_{33}-\left(b\ddot{b}\right) 
h_{00} \right)-\frac{2}{b}\left( 
\ddot{b}+\frac{\dot{a}\dot{b}}{a}+\frac{\dot{b}\dot{c}}{c}  \right) h_{22}=0,}\\
	{\Box h_{33}+2\left(\left(\frac{c\dot{a}\dot{c}}{b^3}\right) 
h_{11}+\left(\frac{c\dot{b}\dot{c}}{b^3}\right) h_{22}-\left(c\ddot{c}\right) 
h_{00} \right)-\frac{2}{c}\left( 
\ddot{c}+\frac{\dot{a}\dot{c}}{a}+\frac{\dot{c}\dot{b}}{b}  \right) h_{33}=0,}\\
	{\Box h_{01}-2\left(\frac{\ddot{a}}{a} \right) h_{10} 
-\frac{1}{a}\left(\ddot{a}+\frac{\dot{a}\dot{b}}{b}+\frac{\dot{a}\dot{c}}{c} 
\right)h_{10}-\left(\frac{\ddot{a}}{a}+\frac{\ddot{b}}{b}+\frac{\ddot{c}}{c}  
\right) h_{01}=0,}\\
	{\Box h_{02}-2\left(\frac{\ddot{b}}{b} \right) h_{20} 
-\frac{1}{b}\left(\ddot{b}+\frac{\dot{a}\dot{b}}{a}+\frac{\dot{b}\dot{c}}{c} 
\right)h_{20}-\left( \frac{\ddot{a}}{a}+\frac{\ddot{b}}{b}+\frac{\ddot{c}}{c}  
\right) h_{02}=0,}\\
	{ \Box h_{03}-2\left(\frac{\ddot{c}}{c} \right) 
h_{30}-\frac{1}{c}\left(\ddot{c}+\frac{\dot{c}\dot{a}}{a}+\frac{\dot{c}\dot{b}}{
b} \right)h_{30} -\left( 
\frac{\ddot{a}}{a}+\frac{\ddot{b}}{b}+\frac{\ddot{c}}{c}  \right) h_{03}=0,}\\
	{\Box h_{12}-2\left(\frac{\dot{a}}{a}\frac{\dot{b}}{b} \right) 
h_{21}-\frac{1}{a}\left( 
\ddot{a}+\frac{\dot{a}\dot{b}}{b}+\frac{\dot{a}\dot{c}}{c} \right) 
h_{12}-\frac{1}{b}\left( 
\ddot{b}+\frac{\dot{a}\dot{b}}{a}+\frac{\dot{b}\dot{c}}{c} \right)h_{21}=0,}\\
	{\Box h_{13}-2\left(\frac{\dot{a}}{a}\frac{\dot{c}}{c} \right) 
h_{31}-\frac{1}{a}\left( 
\ddot{a}+\frac{\dot{a}\dot{b}}{b}+\frac{\dot{a}\dot{c}}{c} \right) 
h_{13}-\frac{1}{c}\left( 
\ddot{c}+\frac{\dot{a}\dot{c}}{a}+\frac{\dot{b}\dot{c}}{b} \right)h_{31}=0,}\\
	{\Box h_{23}-2\left(\frac{\dot{b}}{b}\frac{\dot{c}}{c} \right) 
h_{32}-\frac{1}{b}\left( 
\ddot{b}+\frac{\dot{a}\dot{b}}{a}+\frac{\dot{b}\dot{c}}{c} \right) 
h_{23}-\frac{1}{c}\left( 
\ddot{c}+\frac{\dot{a}\dot{c}}{a}+\frac{\dot{b}\dot{c}}{b} \right)h_{32}=0} ~.
\end{eqnarray}
%\end{widetext}
%
Here $x^{1}, x^{2}, x^{3}$ correspond to $x, y, z$ respectively. One can check the consistency of these equations. 
In particular, by choosing the special case of $a(t)=b(t)=c(t)$, one can see that the equation 
system matches with that for the spatially flat FRW metric.

\subsection{Solutions to the gravitational wave equations}

To solve the wave equations for the anisotropic universe, we shall follow the 
same techniques which we have used earlier. We shall use the gauge condition to 
reduce the number of the independent equations. Here we choose a coordinate 
system, in which the wave propagates in $x$ direction. So according to TTS 
gauge, $h_{0\alpha}$ and $h_{1\alpha}$ (for all $\alpha$) components will be 
identically  zero. Hence, we have only two non-zero independent components of 
the metric perturbations $h_{\alpha \beta}$, viz. $h_{22}$ and $h_{23}$. So, we 
are left with two wave equations

%\begin{widetext}
\begin{eqnarray}
{\Box h_{22}+2\left(\left(\frac{b\dot{a}\dot{b}}{b^3}\right) 
h_{11}+\left(\frac{b\dot{b}\dot{c}}{c^3}\right) h_{33}-\left(b\ddot{b}\right) 
h_{00} \right)-\frac{2}{b}\left( 
\ddot{b}+\frac{\dot{a}\dot{b}}{a}+\frac{\dot{b}\dot{c}}{c}  \right) h_{22}=0,}\\
{\Box h_{23}-2\left(\frac{\dot{b}}{b}\frac{\dot{c}}{c} \right) 
h_{32}-\frac{1}{b}\left( 
\ddot{b}+\frac{\dot{a}\dot{b}}{a}+\frac{\dot{b}\dot{c}}{c} \right) 
h_{23}-\frac{1}{c}\left( 
\ddot{c}+\frac{\dot{a}\dot{c}}{a}+\frac{\dot{b}\dot{c}}{b} \right)h_{32}=0,}
\end{eqnarray}
%\end{widetext}

To integrate these equations, at first we shall expand the d'Alembertian. To 
simplify the calculations, we shall take that the 
perturbation components are only function of $x$ and $t$.  We express two 
independent components of the metric perturbation as $h_{22}=\phi(t,x)$ 
and $h_{23}=\psi(t,x)$. The d'Alembertian simplifies 
to the following form,
\begin{eqnarray} \label{bianchigravwave22}
\Box h_{22} = \Box \phi = 
-\ddot{\phi}-\left(\frac{\dot{a}}{a}+\frac{\dot{b}}{b}+\frac{\dot{c}}{c}
\right)\dot{\phi}+\frac{1}{a^2}\phi^{''}  ~,\\
\Box h_{23}=\Box \psi = 
-\ddot{\psi}-\left(\frac{\dot{a}}{a}+\frac{\dot{b}}{b}+\frac{\dot{c}}{c}
\right)\dot{\psi}+\frac{1}{a^2}\psi^{''} ~, \label{bianchigravwave23}
\end{eqnarray}
where a dot represents the time-derivative and the prime represents the 
derivative with respect to $x$-coordinate. For definiteness, we consider 
anisotropic Kasner metric solution \cite{Kasner:1921zz} to describe the 
background spacetime dynamics \cite{Hu:1978td}. The Kasner metric solution 
is an exact solution of Einstein's field equation and is  given by 
\begin{equation}\label{kasner}
 ds^2=-dt^2+t^{2p_1}dx^2+t^{2p_2}dy^2+t^{2p_3}dz^2,
\end{equation}
where constants $p_1,p_2,p_3$ are known as Kasner exponents. The Kasner 
metric represents a spatially flat universe but depending upon the 
values of $(p_1,p_2,p_3)$ the universe may expand or contract at different rates 
in different directions. The Kasner exponents satisfy following two conditions
\begin{equation}
\sum_{i=1}^{3}p_i=1  ~,~ \sum_{i=1}^{3}p^2_i=1  ~.
\end{equation}
In other words, we shall consider $a(t)=t^{p_1}$, $b(t)=t^{p_2}$ and 
$c(t)=t^{p_3}$ in the wave equations for $h_{22}$ and $h_{23}$
(Eqs. \ref{bianchigravwave22}, \ref{bianchigravwave23}).

\subsubsection{Propagation in the plane of symmetry}

Let us consider the Kasner exponents to be 
$(p_1,p_2,p_3)$=$\left(\frac{2}{3},\frac{2}{3},-\frac{1}{3} \right)$, so that 
the universe is expanding at the same rate in the $x$ and $y$ directions 
while it is contracting in the $z$ direction. As a result, there is a plane of 
symmetry in the $x$-$y$ plane. We have already assumed the wave is propagating 
along the $x$ direction. Now using $h_{\alpha 1}=0$ and $h_{\alpha 0}=0$, for 
all $\alpha$, we get $h_{00}=h_{11}=0$. Also the traceless condition gives 
$h_{22}=-h_{33}$. We substitute $a(t)=t^{p_1}$, $b(t)=t^{p_2}$ and 
$c(t)=t^{p_3}$ with 
$(p_1,p_2,p_3)=\left(\frac{2}{3},\frac{2}{3},-\frac{1}{3}\right)$ in the wave 
equation for $h_{22}=\phi(t,x)$. The wave equation for $\phi$ then becomes
\begin{equation}\label{kasnerwave1}
\Box \phi +\left(\frac{4}{9} \right) \phi = 0 ~,
\end{equation}
which yields
\begin{equation}\label{kasnerwave2}
\ddot{\phi}+\left(\frac{1}{t}\right)\dot{\phi}-\left(\frac{1}{t^{\frac{4}{3}}} 
\right)\phi^{''}-\left(\frac{4}{9} \right)\phi = 0 ~.
\end{equation}
By considering $\phi(t,x)=X_5(x)T_5(t)$, the $x$-dependent differential 
equation becomes
\begin{equation}\label{k1space}
 X_5^{''}+{\Omega}^2 X_5=0  ~,
\end{equation}
where ${\Omega}^2$ is the separation constant. The solution is simple,
\begin{equation}\label{k1spacesol}
 X_5(x)= A \cos({\Omega}x)+B \sin({\Omega}x),
\end{equation}
 where $A$ and $B$ are constants of integration. The temporal part of the
differential equation is given by 
\begin{align}\label{k1time}
\ddot{T_5}+\left(\frac{1}{t}\right)\dot{T_5}-\left(\frac{4}{9} 
-\frac{{\Omega}^2}{t^{\frac{4}{3}}} \right)T_5=0  ~.
\end{align}
It appears that the equation (\ref{k1time}) is not known to have any  
analytic solution. 

On the other hand, the wave equation for $h_{23}=\psi(t,x)$ component can be 
expressed as
\begin{equation}
\Box \psi + \left(\frac{4}{9 t^{2}} \right) \psi = 0 ~, 
\end{equation}
which leads to
\begin{equation}
\ddot{\psi}+\left(\frac{1}{t}\right)\dot{\psi}-\left(\frac{1}{t^{\frac{4}{3}}}
\right)\psi^{''} - \left(\frac{4}{9 t^2}\right)\psi = 0 ~.
\end{equation}
We consider $\psi(t,x)=X_6(x)T_6(t)$. As earlier, the $x$-dependent 
differential equations becomes
\begin{equation}\label{k2space}
X_6^{''}+{\Omega}^2 X_6=0 ~, 
\end{equation}
which has the solution $X_6(x)= C\cos({\Omega}x)+D\sin({\Omega}x)$, with $C$ 
and $D$ being constants of integration. The temporal part of differential 
equations is given by
\begin{align} \label{k2time}
\ddot{T_6}+\left(\frac{1}{t}\right)\dot{T_6}-\left(\frac{4}{9 t^2} 
-\frac{{\Omega}^2}{t^{\frac{4}{3}}} \right)T_6 = 0  ~.
\end{align}
The analytical solution to the equation (\ref{k2time}) is given by
\begin{align}
T_6(t)= C_{9} J_2\left(3{\Omega}t^{\frac{1}{3}} \right) + 
C_{10} Y_2\left(3{\Omega}t^{\frac{1}{3}}\right)  ~,
\end{align}
where $C_9$, $C_{10}$ are the constants of integration. Here $J_n(z)$, $Y_n(z)$ 
are the Bessel functions of the first kind and the second kind respectively. 
The parameter $n$ refers to the order of the Bessel functions.

\begin{figure}[H]
		\centering
		\includegraphics[width=0.7\linewidth]{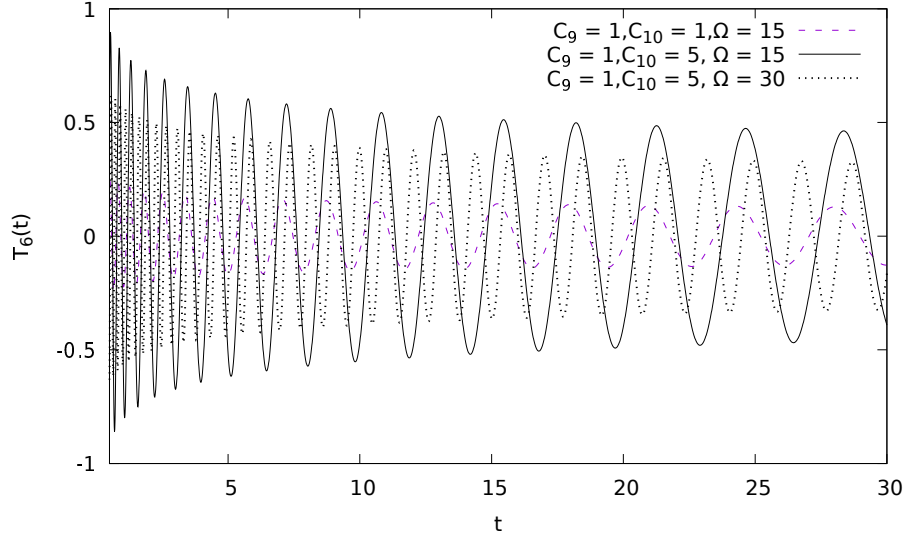}
	\caption{ Plot of evolution of gravitational wave with time in Kasner 
metric where plane of symmetry is in the $x$-$y$ plane.}% The time axis is plotted 
%using arbitrary unit. }
\label{fig:b1inplane}
\end{figure}

We plot $T_6$ for various combinations of the constants of integration and the 
constant of separation in the FIG. \ref{fig:b1inplane}. We can clearly see from 
the plot that the frequency and amplitude are decreasing with time. The rate of 
diminution is different for the different cases. The amplitude is scaled by 
its initial value and the unit of time is $H_0^{-1}$ as $p_1 + p_2 +p_3 =1$ where 
$H = \frac{1}{3} (H_1 + H_2 + H_3)$. Here $H_i$ is the Hubble expansion rate along 
the $i$th spatial direction, e.g., $H_1 = \frac{\dot{a}}{a}$ and so on. The suffix 
$0$ indicates the value of the quantity at $a=b=c=1$ in the present context of scaling. 

\subsubsection{Propagation along the normal to the plane of symmetry}

In this case, we consider the gravitational wave to be propagating along 
the normal to the plane of symmetry. Let us consider 
$(p_1,p_2,p_3)$=$\left(-\frac{1}{3},\frac{2}{3},\frac{2}{3} \right)$, so that 
the universe is expanding at the same rate in $y$ and $z$ directions while 
contracting in $x$ direction. Therefore, there is a plane of symmetry in the 
$y$-$z$ plane and since the gravitational wave is propagating in the 
$x$-direction, this case corresponds to the wave propagating normal to this 
plane. Here again we choose $h_{00}=h_{11}=0$ and the traceless condition gives 
$h_{22}=-h_{33}$. We substitute $a(t)=t^{p_1}$, $b(t)=t^{p_2}$ and 
$c(t)=t^{p_3}$ with the chosen Kasner exponents in the wave equation.
In this case of gravitational wave propagating along the normal to the 
plane of symmetry, the differential equations for $h_{22}=\phi(t,x)$ and 
$h_{23}=\psi(t,x)$ are exactly same and given by
\begin{eqnarray}\label{kasnerwave3}
\ddot{\psi}+\left(\frac{1}{t}\right)\dot{\psi}-\left(\frac{1}{t^{-{\frac{2}{3}}}
}\right)\psi^{''} + \left(\frac{8}{9 t^2} \right)\psi=0.
\end{eqnarray}
Using separation of variables, we consider $\psi(t,x)=X_7(x)T_7(t)$, which 
yields the $x$-dependent differential equation as $X_7^{''}+{\Omega}^2 X_7=0,$ 
where ${\Omega}^2$ is the separation constant. The corresponding solution is 
$X_7(x)= G \cos({\Omega}x)+H \sin({\Omega}x)$ where $G$ and $H$ are constants of 
integration. The $t$-dependent differential equation is given by
\begin{align}\label{k3time}
\ddot{T_7}+\left(\frac{1}{t}\right)\dot{T_7}+\left(\frac{8}{9 t^2} 
+{\Omega}^2 t^{\frac{2}{3}} \right)T_7=0 ~,
\end{align}
which has an analytical solution of the form
%\begin{widetext}
\begin{align}\label{k3}
T_7(t)= J_{-\frac{i}{\sqrt{2}}}\left(\frac{3}{4}\Omega t^{\frac{4}{3}} \right) 
\Gamma\left(1-\frac{i}{\sqrt{2}}\right) C_{11} \nonumber\\
+ J_{\frac{i}{\sqrt{2}}}\left(\frac{3}{4}\Omega t^{\frac{4}{3}} \right) 
\Gamma\left(1+\frac{i}{\sqrt{2}}\right)C_{12}  ~.
\end{align}
%\end{widetext}
Here, $C_{11}$, $C_{12}$ are constants of integration. $J_n(z)$ is the 
Bessel function of first kind where $n$ is its order and $\Gamma(z)$ is the 
Gamma function with argument $z$.

\begin{figure}[H]
		\centering
		\includegraphics[width=0.7\linewidth]{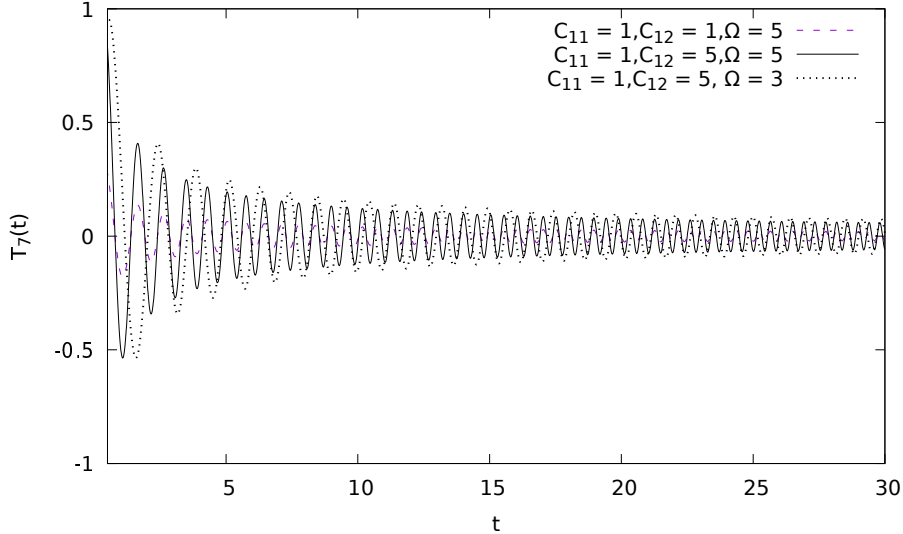}
			\caption{ Plot of evolution of gravitational wave with time in 
Kasner metric where plane of symmetry is $y$-$z$ plane.}% The time axis is plotted 
%using arbitrary unit. }
\label{fig:b1normalplane}
\end{figure}

For various choices of the constants, the plot for $T_7(t)$ is given in 
the FIG. \ref{fig:b1normalplane}. As in the previous case, the frequency and 
amplitude both are decreasing with time.

\section{Discussion}

In this work we present exact solutions of the equation for the 
gravitational waves in some cosmological backgrounds in the linear regime 
of the perturbation where the nonlinear contributions from the perturbations 
are not considered. This linearized wave equation is 
solved for the dust dominated and also the radiation dominated 
spatially flat isotropic universe. As already mentioned in section 2,
we worked in a short wave condition. The idea was not to look at the full 
spectrum of the gravity wave. The interest clearly was to see how the amplitude 
and frequency of the wave are modified in the cosmologial background. Qualitatively, 
other modes, if any, should also undergo similar changes.
Both the amplitude and frequency are found 
to decay with time (figures \ref{fig:dust} and \ref{fig:radflat}). For a 
spatially closed universe, the exact solutions are obtained for two cases. For a 
standard radiation dominated universe, the amplitude and frequency decrease to 
start with, but after reaching a minimum, increases at the same rate (figure 
\ref{fig:radclose}). This apparently intriguing feature is actually quite  
consistent, as the universe recollapses, after reaching a maximum volume, into a 
big crunch in this model. The other example for the closed universe is for the 
so-called K-matter distribution ($P = - \frac{1}{3} \rho$) which gives Milne's 
``coasting'' universe which expands with a zero deceleration parameter $q$. The 
amplitude and frequency both monotonically decay as given in figure 
\ref{fig:coasting}. The temporal parts of the two degrees of freedom (or the two 
modes) of the gravitational wave have the same equation for any given isotropic 
model.

For the anisotropic case, we take up the Kasner metric as the example with 
a plane of symmetry. If the wave propagates in the plane of symmetry, the 
two modes have different equations for the temporal part. For one mode ($h_{23} 
= h_{32}$) the exact solution is obtained, whereas for the other ($h_{22} = - 
h_{33}$) we failed to get an analytic solution. For the wave propagating normal 
to the plane of symmetry, the two modes behaves identically, and the exact 
solution is indeed obtained. The wave form for the exactly solved cases are 
given in figures \ref{fig:b1inplane} and \ref{fig:b1normalplane}. The behaviour 
is similar to the isotropic expansion, a monotonic decay of both the amplitude 
and frequency.

As already mentioned, the exact numerical values for the initial conditions 
that fixes the constants of integration, or the exact values of other 
constants (such as the constant of separation) are not rigorously settled in 
any way for the present work. The focus is more on the exact solutions of the 
wave equations, and it is clearly demonstrated that they can be written, at 
least in some special cases. The examples of the cosmologies chosen are not 
contrived, they are there in the literature and are quite widely used. The time 
axis in the plots are scaled not to any standard unit, but it does not obscure 
the nature of the plots in any way. Each of the plots is a family of curves, 
where each member represents a set of constants. It is quite apparent from all 
the figures that the nature of the evolution is not too sensitive to the choice 
of the constants.

\end{document}